\documentclass[a4,12pt]{article}

\usepackage{amsmath,amssymb}
\usepackage{bm}

\makeatletter

\@addtoreset{equation}{section}
\makeatother

\newcommand{\rap}[2]
{\setbox1=\hbox{#1}%
\setbox2=\hbox to\wd1{\hss #2\hss}%
\mbox{\rlap{\box1}\box2}}

\newcommand{\wt}{\widetilde}

\newcommand{\wh}{\widehat}
\newcommand{\ol}{\overline}

\newcommand{\RR}{\mathbb{R}}
\newcommand{\CC}{\mathbb{C}}

\newcommand{\N}{\mathcal{N}}
\newcommand{\ZZ}{\mathbb{Z}}

\usepackage{version}	
\begin{document}
\begin{titlepage}
\title{
\vspace{-2cm}
\begin{flushright}
\normalsize{
TIT/HEP-662\\
November 2017}
\end{flushright}
       \vspace{1cm}
       Codimension-2 brane solutions of maximal supergravities in 9, 8, and 7 dimensions
       \vspace{1cm}}
\author{
Yosuke Imamura\thanks{E-mail: \tt imamura@phys.titech.ac.jp}$^{~1}$,
Hirotaka Kato\thanks{E-mail: \tt h.kato@th.phys.titech.ac.jp}$^{~1}$
\\[30pt]
{\it $^1$ Department of Physics, Tokyo Institute of Technology,}\\
{\it Tokyo 152-8551, Japan}\\
}

\thispagestyle{empty}

\vspace{0cm}

\maketitle

\begin{abstract}
We construct codimension-2 BPS brane solutions 
in $D=9,8,7$ maximal supergravities by solving Killing spinor equations.
We assume the Poincare invariance along the worldvolume and
vanishing gauge fields, and determine the metric and the scalar fields.
The solution in $D=9$ is essentially the same as the ten-dimensional one,
which is specified by a holomorphic function in the transverse space.
For $D=8$, the solution is specified by two holomorphic functions,
and regarded as $T^2\times T^2$ compactification of F-theory.
For $D=7$, we find that the solution can be interpreted as
M-theory on Calabi-Yau, and under an additional assumption
a solution is specified by two holomorphic functions.
\end{abstract}
\end{titlepage}


\section{Introduction}
An important feature of codimension $2$-branes is 
that they can have non-trivial monodromies \cite{Greene:1989ya}.
Namely, when we move charged objects around
such branes they may get transformed to dual objects.
The element of duality group specifying this duality transformation
is called monodromy associated with the branes.
In the context of brane realization of
field theories 
such monodromy transformations are often interpreted as electric-magnetic
duality,
and in some cases the existence of branes with non-trivial monodromies 
causes emergence of particles with 
mutually non-local charges.
This is a common feature of
some classes of non-Lagrangian theories
such as Argyres-Douglas theories \cite{Argyres:1995jj,Argyres:1995xn}
and 4d ${\cal N}=3$ superconformal theories \cite{Aharony:2015oyb,Garcia-Etxebarria:2015wns,Garcia-Etxebarria:2016erx}.
\footnote{Non-trivial monodromies arise not only in
systems of codimension $2$-branes but also
in more general backgrounds with
non-trivial fundamental groups.
Indeed, the first example of ${\cal N}=3$
theory in \cite{Garcia-Etxebarria:2015wns} is realized by using an orbifold $\CC^3/\ZZ_k$,
which has the fundamental group $\ZZ_k$.
Another example of ${\cal N}=3$ theories
in \cite{Garcia-Etxebarria:2016erx} can be regarded as a realization with
codimension $2$-branes.}
This fact motivates us to investigate
codimension $2$-branes.

In the context of string/M-theory a maximal supergravity is
obtained by torus compactification of ten- or eleven-dimensional theory.
The U-duality group is generated by geometric coordinate changes
of the torus and duality transformations.
Half BPS branes in eleven- or ten-dimensions descent to various sorts of codimension-2 branes in
lower dimensions by dimensional reduction and duality transformations\cite{Elitzur:1997zn,Obers:1998fb,deBoer:2010ud}.

F- and M-theories are useful to describe codimension-2 branes.
For example, $7$-branes in type IIB string theory
are described as purely geometric objects in
the context of F-theory.
We can also realize various branes in lower dimensions
by considersing M- and F-theory in different purely geometric backgrounds
in which fields except the metric are
vanishing or constant.
However, there may be branes which does not have geometric description in
M- or F-theory.
Such branes have not been investigated in detail,
and their realization and classification
may give new insight for strongly coupled field theories.
A purpose of this paper is to search for
such branes in the case of codimension-$2$ BPS branes.


Actually, construction of such branes is quite restricted
as is argued in \cite{Kumar:1996zx}.
Some examples given in \cite{Kumar:1996zx}
have the non-compact dimensions less than four,
and it seems difficult to give examples in higher dimensions.
We show that this is actually the case
for codimension-$2$ BPS brane solutions by explicitly solving Killing spinor equations.
Namely, BPS solutions of codimension-$2$ branes always
have geometric realization in M/F-theory
for $D=9,8,7$.

Maximal supergravities in various dimensions
have common structure.
The scalar manifolds of these theories have the form $G/H$,
where $G$ is the classical global symmetry 
and
$H$ is the local symmetry group,
which is the maximal compact subgroup of $G$.
See Table \ref{groups} for $G$ and $H$ 
in dimensions $D=10,\ldots,4$ \cite{Hull:1994ys}.
\begin{table}[htb]
\caption{The global symmetry group $G$, the duality group $G_{\ZZ}$, and the local symmetry group $H$ in maximal supergravities}\label{groups}
\centering
\begin{tabular}{cccc}
\hline
\hline
 dim & $G$  & $G_\ZZ$ & $H$  \\
\hline
10(A) & $SO(1,1)/\mathbb{Z}_2$  & $1$ & $1$ \\
10(B) & $SL(2,\mathbb{R})$ & $SL(2,\mathbb{Z})$ & $SO(2)$ \\
9 & $SL(2,\mathbb{R})\times O(1,1)$ & $SL(2,\mathbb{Z})\times\mathbb{Z}_2$ & $SO(2)$\\
8 & $SL(3,\mathbb{R})\times SL(2,\mathbb{R})$ & $SL(3,\mathbb{Z})\times SL(2,\mathbb{Z})$ & $SO(3)\times SO(2)$ \\
7 & $SL(5,\mathbb{R})$ & $SL(5,\mathbb{Z})$ & $SO(5)$ \\ 
6 & $O(5,5)$ & $O(5,5;\mathbb{Z})$ & $SO(5)\times SO(5)$ \\ 
5 & $E_{6(6)}$ & $E_{6(6)}(\mathbb{Z})$ & $USp(8)$ \\ 
4 & $E_{7(7)}$ & $E_{7(7)}(\mathbb{Z})$ & $SU(8)$ \\ \hline
\end{tabular}
\end{table}
The scalar fields are coordinates of this manifold,
and represented as a matrix $L\in G$ with 
left action of $G$ and right action of $H$.
The U-duality group $G_\ZZ$ is the integral form of $G$.
For the theories in seven or higher dimensions
$G$ are all $SL$ type and
$H$ are all $SO$ type,
and they can be dealt in similar ways.
In this paper we investigate these theories,
and theories in $D\leq 6$ are left for future work.

A maximal supergravity contains
\begin{itemize}
\item
the vielbein $e_M^{\wh M}$
\item
scalar fields $L^\alpha{}_i$
\item
gravitino $\psi_M$
\item
dilatino $\lambda_i$
\end{itemize}
and anti-symmetric tensor fields of
different ranks,
which are not relevant to our analysis in this paper.

We use the following indices:
\begin{itemize}
\item $M,N,\ldots$ : global coordinates
\item $\wh M,\wh N,\ldots$ : local Lorentz
\item $\alpha,\beta,\ldots$ : $SL(m)$ fundamental representation
\item $i,j,\ldots$ : $H=SO(n)$ vector.
\end{itemize}

The scalar fields appear in the action and the supersymmetry transformation laws
through 1-form fields $P$ and $Q$, which are defined as the traceless symmetric and anti-symmetric parts
of the Maurer-Cartan form:
\begin{align}
P_{ij}=(L^{-1}dL)_{(ij)},\quad
Q_{ij}=(L^{-1}dL)_{[ij]}. \label{PQdef}
\end{align}
Under $H$ transformation $P$ transforms homogeneously as
the symmetric matrix representation of $H$,
while $Q$ transforms inhomogeneously and plays the role of
$H$-connection.

To obtain BPS solutions we solve the Killing spinor equations for the
gravitino $\psi_M$ and dilatino $\lambda_i$.
In the next section we first look at 10d case to explain basic prescription to
solve the Killing spinor equations and then we move on to lower dimensional cases.

\section{Solving Killing spinor equations}
\subsection{$D=10$}\label{7brane_in_10dim}
Let us consider BPS solutions in type IIB supergravity.
Such solutions have been well investigated\cite{Greene:1989ya} and 
7-branes are classified by Kodaira classification\cite{Kodaira}.
Various 4d $\N=2$ supersymmetric theories are 
realized on D3-branes probing these solutions\cite{Sen:1996vd,Dasgupta:1996ij,Gaberdiel:1997ud}. 
A purpose of this subsection is
to review how we can obtain BPS solutions
in ten dimensions by solving the Killing spinor equations.
The derivations in lower dimensions are parallel.

The classical global symmetry of type IIB supergravity is $G=SL(2,\RR)$
and the local R-symmetry group is $H=SO(2)_R$.
Namely, the scalar manifold is locally the two-dimensional homogeneous space
$SL(2,\RR)/SO(2)_R$.
When we discuss the global structure, we also need to take account of
the duality group $G_{\ZZ}=SL(2,\ZZ)$.
Quantum numbers of scalar and spinor fields
in type IIB supergravity \cite{SchwarzIIB} are summarized
in Table \ref{IIB.tbl}.
\begin{table}[htb]
\caption{Quantum numbers of scalar and spinor fields in type IIB supergravity}\label{IIB.tbl}
\centering
\begin{tabular}{ccc}
\hline
\hline
& $G=SL(2,\RR)$ & $H=SO(2)_R$ \\
\hline
$L^\alpha{}_i$ & $\bm2$ & $\pm1$ \\
$\psi_M$ & $\bm1$ & $\pm\frac{1}{2}$ \\
$\lambda_i$ & $\bm1$ & $\pm\frac{3}{2}$ \\
$\epsilon$ & $\bm1$ & $\pm\frac{1}{2}$ \\
\hline
\end{tabular}
\end{table}
The gravitino field $\psi_M$ belongs to the spinor representation of $H$.
Namely, $\psi_M$ has the spacetime vector index $M$ and an $SO(2)_R$ spinor index
which is implicit.
The dilatino field $\lambda_i$ has 
the $SO(2)_R$ vector index $i$ and 
an implicit $SO(2)_R$ spinor index.
It satisfies the $\rho$-traceless condition
\begin{align}
\rho_i\lambda^i=0,
\end{align}
where $\rho_i$ are Dirac matrices associated with the orthogonal group $H=SO(2)_R$.
See appendix for our notation.
This condition removes components carrying $SO(2)_R$ charge $\pm1/2$ from $\lambda^i$,
and the remaining components in $\lambda^i$ carry $SO(2)_R$ charge $\pm3/2$ 
as is shown in Table \ref{IIB.tbl}.

Due to the existence of the self-dual $4$-form field
it is difficult to write down the full Lagrangian of the
type IIB supergravity.
However, it is easy to give
the Lagrangian of the subsector which is relevant to us.
If we assume the vanishing anti-symmetric tensor fields
the equations of motion for the remaining fields are obtained from the
Lagrangian
\begin{align}
{\cal L}
&=\frac{e}{4}R
+\frac{e}{2}(\psi_M\Gamma^{MNP}D_N\psi_P)
\nonumber\\&
-\frac{e}{4}(P_M{}^{ij})^2
+\frac{e}{2}(\lambda_i\Gamma^ND_N\lambda_i)
+\frac{e}{2}P_M{}^{ij}(\psi_N\Gamma^M\Gamma^N\Gamma_i\lambda_j),
\end{align}
up to higher order fermion terms.

The Killing spinor equations are
\begin{align}
0=\delta\psi_M
&=D_M\epsilon,
\label{iibdpsi}\\
0=\delta\lambda^i
&=P_M^{ij}\Gamma^M\rho_j\epsilon,
\label{iibdlambda}
\end{align}
where $D_M$ is the covariant derivative defined with the spin connection $\omega$
and the $SO(2)_R$ connection $Q$:
\begin{align}
D_M\epsilon=
\left(\partial_M+\frac{1}{4}\omega_{M\wh P\wh Q}\Gamma^{\wh P\wh Q}
+\frac{1}{4}Q_{Mij}\rho^{ij}\right)\epsilon.
\end{align}
We are interested in codimension-2 brane solutions.
Let us assume the solution has the eight dimensional Poincare invariance along the eight longitudinal directions.
We use $x^\mu$ ($\mu=0,1,\ldots,7$) and $x^m$ ($m=8,9$) for longitudinal and transverse coordinates, respectively.
We take the ansatz
\begin{align}
ds^2=f^2(x^m)\eta_{\mu\nu}dx^\mu dx^\nu
+g^2(x^m)dx^m dx^m,
\end{align}
for the metric and
\begin{align}
L^\alpha{}_i=L^\alpha{}_i(x^m)
\end{align}
for the scalar fields.
We introduce the local frame so that the vielbein has the
diagonal components
\begin{align}
e^{\wh\mu}=f(x^m)\delta^{\wh\mu}_\mu dx^\mu,\quad
e^{a}=g(x^m)\delta^a_m dx^m.
\label{commonans}
\end{align}
Because we are interested in the rigid supersymmetry
on the branes we assume the supersymmetry parameter $\epsilon$ depends only on the
transverse coordinates:
\begin{align}
\epsilon=\epsilon(x^m).
\end{align}

Because $L^\alpha{}_i$ is independent of the longitudinal coordinates $x^\mu$,
the longitudinal components of $Q$ vanish.
For the longitudinal components of the Killing spinor equation
(\ref{iibdpsi})
\begin{align}
\delta\psi_\mu
=D_\mu\epsilon
=\left(\partial_\mu-\frac{1}{2g}(\partial_mf)\Gamma_{m\wh\mu}\right)\epsilon
=0
\end{align}
to have
non-trivial solutions the function $f$ must be constant,
and without loss of generality
we can set $f=1$.

The covariant derivative in
the transverse components of 
(\ref{iibdpsi}) include
the connection of $SO(2)_{89}$,
the rotation in the $8$-$9$ plane,
and that of $H=SO(2)_R$:
\begin{align}
D_m\epsilon = 
\left(\partial_m+\frac{1}{2}\omega_{m89}\Gamma^{89}
+\frac{1}{2}Q_{m12}\rho^{12}\right)\epsilon=0.
\end{align}
For the existence of non-vanishing solutions,
the action of two connections on some components
of $\epsilon$ must be pure gauge.
To study this condition, it is convenient
to
decompose the parameter $\epsilon$ into four parts
$\epsilon_{s,r}$ according to $SO(2)_{89}$ and $SO(2)_R$ charges
so that
\begin{align}
\frac{1}{2}\Gamma_{89}\epsilon_{s,r}=is\epsilon_{s,r},\quad
\frac{1}{2}\rho_{12}\epsilon_{s,r}=ir\epsilon_{s,r},
\end{align}
where both the indices $s$ and $r$ 
take values in $\{+\frac{1}{2},-\frac{1}{2}\}$. 
We also decompose $\lambda^i$ in the same way into $\lambda^i_{sr}$.
For distinction we use $s=\{\uparrow,\downarrow\}$ for $SO(2)_{89}$ 
and $r=\{+,-\}$ for $SO(2)_R$. 
We also introduce $i=\{\oplus,\ominus\}$ for complex basis of $SO(2)_R$ vectors, 
which carry $SO(2)_R$ charge $\pm 1$. 
See appendix for detail.

Let us require the solution to be half BPS.
Without loosing generality we can assume
that $\epsilon_{\uparrow +}$ and its Majorana conjugate $\epsilon_{\downarrow -}$
correspond to the unbroken supersymmetries.
The other components are set to be zero:
$\epsilon_{\downarrow+}=\epsilon_{\uparrow-}=0$.
Then, the non-vanishing components of $\delta\lambda$
are
\begin{align}
\delta\lambda^i_{\downarrow-}=P_{z^*}^{i\oplus}\epsilon_{\uparrow +},
\label{dl10}
\end{align}
and its complex conjugate.

Before proceeding, it would be instructive to check the consistency of the
quantum numbers in (\ref{dl10}).
Let us first consider the $SO(2)_{89}$ quantum numbers.
The left hand side has the lower index $\downarrow$.
This means the component carries $SO(2)_{89}$ charge (spin) $-1/2$.
On the right hand side, the parameter $\epsilon$ has lower index $\uparrow$
which means $SO(2)_{89}$ spin $+1/2$.
In addition, $P$ has lower index $z^*$
and this component carries $SO(2)_{89}$ spin $-1$.
Therefore, both left and right hand sides carry
the same $SO(2)_{89}$ spin $-1/2$.
The coincidence of the $SO(2)_R$ charge can be confirmed in a similar way.
The index $i$ is common for left and right hand sides
and thus let us focus on the other indices.
On the left hand side we have lower $-$ index
and this means it carries $SO(2)_R$ charge $-1/2$.
On the right hand side
there are the upper $\oplus$ index on $P$ and
the lower $+$ index on $\epsilon$,
which carry $SO(2)_R$ charges $-1$ and $+1/2$, respectively.
Therefore, the left and right hand sides carry the same $SO(2)_R$ charge
$-1/2$.
The charge counting we have just explained is quite useful
when we extract condition imposed on $P$ from Killing spinor equations
associated with dilatino fields in different dimensions.

The vanishing of (\ref{dl10}) mean
\begin{align}
P_{z^*}^{\oplus\oplus}=0.
\label{pzpp}
\end{align}
($P_{z^*}^{\ominus\oplus}$ is identically zero due to the traceless condition.)
We want to solve this with respect to the scalar fields $L^\alpha{}_i$.
For this purpose it is convenient to gauge fix the local $SO(2)_R$ symmetry
so that the matrix $L$ is given by
\begin{align}
L=
K(\tau),
\label{tendimL}
\end{align}
where $K(\tau)$ for a complex number $\tau$ in the upper half plane is
the following $2\times2$ matrix:
\begin{align}
K(\tau)=\frac{1}{\sqrt{\tau_2}}
\left(\begin{array}{cc}
1 & 0 \\
\tau_1 & \tau_2
\end{array}\right),\quad
\tau\equiv\tau_1+i\tau_2\in H_+. 
\label{ktaudef}
\end{align}
Then $P$ and $Q$ have the components
\begin{align}
P^{ij}=\frac{1}{2\tau_2}\left(\begin{array}{cc}
-d\tau_2 & d\tau_1 \\
d\tau_1 & d\tau_2
\end{array}\right),\quad
Q^{ij}=\frac{d\tau_1}{2\tau_2}\left(\begin{array}{cc}
0 & -1 \\
1 & 0
\end{array}\right).
\end{align}
In this gauge the equation (\ref{pzpp}) gives
\begin{align}
P_{z^*}^{\oplus\oplus}=\frac{i}{2\tau_2}\partial_{z^*}\tau=0.
\end{align}
Namely, $\tau$ must be a holomorphic function of $z$.

Now let us turn to the equation $\delta\psi_m=D_m\epsilon=0$. 
The components including the non-vanishing parameters $\epsilon_{\uparrow +}$ and $\epsilon_{\downarrow -}$ 
are 
\begin{align}
D_m\epsilon_{\uparrow +}
= \left(\partial_m+\frac{i}{2}(\omega_{m89}+Q_{m12})\right)\epsilon_{\uparrow +}
=0
\label{dme0in10}
\end{align}
and its complex conjugation.
For (\ref{dme0in10})
to have solutions with $\epsilon_{\uparrow +}\neq0$,
the net connection $\omega_{89}+Q_{12}$ must be pure gauge, and
we can take the gauge with $\omega_{89}+Q_{12}=0$.
The explicit form of the spin connection and the $SO(2)_R$ connection are
\begin{align}
\omega_{89}
=i\frac{\partial g}{g}dz
-i\frac{\ol\partial g}{g}dz^*,\quad
Q_{12}=-\frac{d\tau_1}{2\tau_2}
=-i\frac{\partial\tau_2}{2\tau_2}dz
+i\frac{\ol\partial\tau_2}{2\tau_2}dz^*.
\end{align}
where we used holomorphy of $\tau$ in the last equality.
From $\omega_{89}+Q_{12}=0$ we obtain
\begin{align}
\frac{dg}{g}=\frac{d\tau_2}{2\tau_2}
\end{align}
and this is solved by
\begin{align}
g=c\sqrt{\tau_2},
\end{align}
where $c$ is an arbitrary real positive constant,
which can be absorbed by the coordinate change $cx^m\rightarrow x^m$.

The solution is summarized as follows.
\begin{align}
L^\alpha{}_i
&=K(\tau),\\
ds^2&=\eta_{\mu\nu}dx^\mu dx^\nu+\tau_2dx^m dx^m,\\
\tau(z)&=\tau_1+i\tau_2,\quad
\tau_2>0.
\end{align}
This solution is specified by the single holomorphic function $\tau(z)$.

The imaginary part of $\tau(z)$ must be positive,
and no globally defined holomorphic function satisfy
this condition unless $\tau(z)$ is a constant.
For non-trivial solution $\tau(z)$ must be given as
multi-valued solution with singularities.
These singularities are regarded as branes, and
the monodromies associated with the multi-valueness
specify the charges of the branes.
It is well-known that these singularities are classified by
Kodaira classification, and
we do not give detailed explanation about it.

In the following we will construct solutions in lower dimensions,
and find that they are also described by holomorphic functions
with positive imaginary part.
Because the classification of the singularity can be done in
a similar way to the ten-dimensional case, and it is well studied,
we only focus on the local structure of solutions.

\subsection{$D=9$}

Let us start the analysis in lower dimensions following 
the prescription in the last subsection. 
The scalar and spinor fields
in the nine-dimensional ${\cal N}=2$ supergravity \cite{Nishino:2002zi}
are summarized in Table \ref{9d.tbl}.
\begin{table}[htb]
\caption{The quantum numbers of scalar and spinor fields in the nine-dimensional ${\cal N}=2$ supergravity}\label{9d.tbl}
\centering
\begin{tabular}{ccc}
\hline
\hline
& $SL(2)$ & $SO(2)$ \\
\hline
$L^\alpha{}_i$ & $\bm2$ & $\pm1$ \\
$\varphi$ & $\bm1$ & $0$ \\
$\psi_M$ & $\bm1$ & $\pm\frac{1}{2}$ \\
$\lambda_i$ & $\bm1$ & $\pm\frac{3}{2}$ \\
$\wt\lambda$ & $\bm1$ & $\pm\frac{1}{2}$ \\
$\epsilon$ & $\bm1$ & $\pm\frac{1}{2}$ \\
\hline
\end{tabular}
\end{table}
The fields $\lambda_i$ are subject to the gamma-traceless condition $\rho_i\lambda_i=0$.
The Lagrangian is
\begin{align}
 {\cal L}
 =& -\frac{e}{4}R
 - \frac{i}{2}e(\psi_L\Gamma^{LMN}D_M\psi_N)
 \nonumber\\&
 + \frac{e}{4}(P_{M ij})^2
 + \frac{i}{2}e(\lambda_i\Gamma^M D_M\lambda_i) 
 + \frac{i}{2}e(\psi_M\rho_i\Gamma^N\Gamma^M\lambda_j)P_{Nij} 
\nonumber\\&
 + \frac{e}{2}(\partial_M\varphi)^2
 + \frac{i}{2}e(\wt\lambda\Gamma^M D_M\wt\lambda)
 + \frac{i}{\sqrt{2}}e(\psi_M\Gamma^N\Gamma^M\wt\lambda_j)\partial_N\varphi
 + \ldots ,
\end{align}
where the dots represent terms with gauge fields and four-fermi terms,
which play no role in the following analysis.
The supersymmetry transformation rules for the spinor fields are
\begin{align}
 \delta\psi_M &= D_M\epsilon,\\
 \delta\lambda_i &= \frac{1}{2}P_{M ij}\Gamma^M\rho_j\epsilon ,\\
 \delta\wt\lambda &= \frac{1}{\sqrt{2}}D_M\varphi\Gamma^M\epsilon .
\end{align}

We want to obtain codimension $2$ brane solutions by solving
the Killing spinor equations.
We use $x^\mu$ ($\mu=0,1,\ldots,6$) and $x^m$ ($m=7,8$) for longitudinal and transverse coordinates, respectively.
We take the ansatz
\begin{align}
 L^\alpha{}_i = L^\alpha{}_i(x^m),\quad
 \varphi = \varphi(x^m), \quad
 e^{\hat{\mu}} = \delta^{\hat{\mu}}_\mu dx^\mu, \quad
 e^a = g(x^m)\delta^a_m dx^m, \quad
 \epsilon = \epsilon(x^m)
\end{align}
In fact, the solution is almost the same as that of type IIB case.
Although we have extra fields $\varphi$ and $\wt\lambda$ compared to the ten-dimensional 
case,
the condition $\delta\wt\lambda = 0$ forces $\varphi$ to be constant;
\begin{align}
0=\delta\wt\lambda
&= \frac{1}{\sqrt{2}}\partial_m\varphi\Gamma^m\epsilon
\quad
\rightarrow
\quad
 \partial_m\varphi = 0.
\end{align}
Therefore, we can forget about $\wt\lambda$ and $\varphi$, and remaining fields
give the set of equations identical to the ten-dimensional case.
After some gauge choices
the general solution is given by
\begin{align}
\varphi&=\mathrm{const}\\
L^\alpha{}_i&=K(\tau),\quad
\tau=\tau_1 + i\tau_2 : \mbox{holomorphic function}\\
ds^2&=\eta_{\mu\nu}dx^\mu dx^\nu+\tau_2dx^m dx^m.
\end{align}
A solution is specified by a single holomorphic
function $\tau(z)$ and a constant vacuum expectation value of
$\varphi$.
Cosimenison $2$ brane solutions appear as singularities of the function $\tau(z)$.

\subsection{$D=8$}\label{5branein8d}
The scalar and fermion fields
in $8$d maximal supergravity \cite{Salam:1984ft} are shown in in Table \ref{8d.tbl}.
\begin{table}[htb]
\caption{Quantum numbers of scalar and spinor fields in 8d maximal
supergravity.
The $SO(2)_R$ charge of each component of a spinor is proportional
to the chirality.
}\label{8d.tbl}
\centering
\begin{tabular}{ccc}
\hline
\hline
& $SO(3)_R$ & $SO(2)_R$ \\
\hline
$\wt L^{\wt\alpha}{}_{\wt i}$ & $\bm3$ & $0$ \\
$L^\alpha{}_i$ & $\bm1$ & $\pm 1$ \\
$\psi_M$ & $\bm2$ & $\frac{1}{2}\Gamma_9$ \\
$\lambda_i$ & $\bm2$ & $\frac{3}{2}\Gamma_9$ \\
$\wt\lambda_{\wt i}$ & $\bm4$ & $-\frac{1}{2}\Gamma_9$ \\
\hline
\end{tabular}
\end{table}
Classical $p$-brane solutions with $p=0,1,3,4$ are given in \cite{Lu:1998sx}.
Half BPS solutions of 10 dimensional supergravity given in \cite{Kimura:2014wga} 
can be regarded as codim-2 branes in 8 dimensional supergravity.
In the following we construct general 5-brane solutions 
without assuming 10 dimensional supergravity description.

The scalar manifold of 
the eight dimensional maximal supergravity
is the direct product of two homogeneous spaces: $SL(2,\ZZ)/SO(2)_R\times SL(3,\ZZ)/SO(3)_R$.
Each factor can be interpreted geometrically in an appropriate
duality frame.
The $SL(2,\ZZ)/SO(2)_R$ becomes manifest when we regard the theory as
$T^2$ compactification of type IIB theory,
while $SL(3,\ZZ)/SO(3)_R$ can be regarded as the moduli space
associated with $T^3$ compactification of M-theory.
The S-duality group in the type IIB picture
is a subgroup of $SL(3,\ZZ)$.

For each factor of the R-symmetry group $SO(2)_R\times SO(3)_R$
there is associated dilatino field.
We denote fields associated with $SO(2)_R$ and $SO(3)_R$
by $\lambda^i$ and $\wt\lambda^{\wt i}$, respectively.
All fermion fields have implicit spinor indices for
all $SO(1,7)$, $SO(2)_R$, and $SO(3)_R$.
In addition, $\lambda$ and $\wt\lambda$ have $SO(2)$ and $SO(3)$ vector indices, respectively,
and they satisfy the traceless conditions
$\rho_i\lambda^i=0$ and $\wt\rho_{\wt i}\wt\lambda^{\wt i}=0$.
Namely, $\lambda$ and $\wt\lambda$ belong to $2_{\pm\frac{3}{2}}$ and $4_{\pm\frac{1}{2}}$, respectively,
of $SO(2)_R\times SO(3)_R$.

The Lagrangian is
\begin{align}
{\cal L}
&=\frac{e}{4}R
+\frac{e}{2}(\psi_M\Gamma^{MNP}D_N\psi_P)
\nonumber\\&
-\frac{e}{4}(P_M{}^{ij})^2
+\frac{e}{2}(\lambda_i\Gamma^ND_N\lambda_i)
+\frac{e}{2}P_M{}^{ij}(\psi_N\Gamma^M\Gamma^N\rho_i\lambda_j)
\nonumber\\&
-\frac{e}{4}(\wt P_M{}^{\wt i\wt j})^2
-\frac{e}{2}(\wt\lambda_{\wt i}\Gamma^ND_N\wt\lambda_{\wt i})
+i\frac{e}{2}\wt P_M{}^{\wt i\wt j}(\psi_N\Gamma^M\Gamma^N\Gamma_{\wt i}\wt\lambda_{\wt j}) 
+ \ldots
\end{align}
where the dots represent four-fermion terms and terms with gauge fields.
The supersymmetry transformation laws of fermions are
\begin{align}
\delta\psi_M&=D_M\epsilon,\\
\delta\lambda^i&=\frac{1}{2}P_M{}^{ij}\Gamma^M\rho_j\epsilon,\\
\delta\wt\lambda^{\wt i}&=\frac{i}{2}\wt P_M{}^{\wt i\wt j}\Gamma^M\wt\rho_{\wt j}\epsilon.
\end{align}

We are interested in codimension $2$-brane solutions and
we use $x^\mu$ ($\mu=0,1,\ldots,5$) and $x^m$ ($m=6,7$) for longitudinal and transverse coordinates, respectively.
We take the following ansatz:
\begin{align}
L^\alpha{}_i=L^\alpha{}_i(x^m),\quad
\wt L^{\wt\alpha}{}_{\wt i}=\wt L^{\wt\alpha}{}_{\wt i}(x^m),\quad
e^{\hat{\mu}}=\delta^{\hat{\mu}}_\mu dx^\mu,\quad
e^a=g(x^m)\delta^a_mdx^m,\quad
\epsilon=\epsilon(x^m)
\end{align}

The covariant derivative $D_M\epsilon$ contains three connections
$\omega$, $Q$, and $\wt Q$, 
corresponding to $SO(2)_{67}$, $SO(2)_R$, and $SO(3)_R$, respectively.
For the existence of non-trivial solution to $\delta\psi_m=0$,
the actions of three connections to some components of $\epsilon$ must 
be pure gauge.
For this to be the case, non-vanishing components of $SO(3)_R$ connection $\wt Q$ should be
in a certain $SO(2)$ subgroup of $SO(3)_R$.
We can take the gauge such that
it is rotation of $\wt 1 \wt 2$ plane and 
\begin{align}
\wt Q^{\wt i\wt3}=0.
\label{qi3}
\end{align}
After taking this gauge, we have three $SO(2)$ connections
$\omega_{67}$, $Q_{12}$ and $\wt Q_{\wt 1\wt 2}$.
As in the $10$ dimensional case
it is convenient to divide the parameter $\epsilon$ into components
$\epsilon_{sr\wt r}$ so that
\begin{align}
\frac{1}{2}\Gamma_{67}\epsilon_{sr\wt r}=is\epsilon_{sr\wt r},\quad
\frac{1}{2}\rho_{12}\epsilon_{sr\wt r}=ir\epsilon_{sr\wt r},\quad
\frac{1}{2}\wt\rho_{12}\epsilon_{sr\wt r}=i\wt r\epsilon_{sr\wt r},
\end{align}
where all of $s$, $r$, and $\wt r$ take values in $\{+\frac{1}{2},-\frac{1}{2}\}$.
For distinction we introduce the notation
$s\in\{\uparrow,\downarrow\}$ for $SO(2)_{67}$,
$r\in\{+,-\}$ for $SO(2)_R$,
and $\wt r\in\{\wt +,\wt -\}$ for $SO(3)_R$.
We also introduce
$\{\oplus,\ominus\}$ for the complex basis of $SO(2)_R$ vector and
$\{\wt\oplus,\wt\ominus,\wt 3\}$ for the basis of $SO(3)_R$ vector 
that diagonalize $SO(2)_{\wt 1\wt 2}$.

The 6d chirality of $\epsilon_{sr\wt r}$
is given by $s$ and $\wt r$ as
\begin{align}
\gamma_7\epsilon_{sr\wt r} = \mathrm{sign}(s\wt r)\epsilon_{sr\wt r}.\label{6dchirality}
\end{align}

We want to consider solution in which some of $\epsilon_{sr\wt r}$ are preserved.
Without loss of generality, we can suppose that
$\epsilon_{\uparrow+\wt+}$ and its complex conjugate
$\epsilon_{\downarrow-\wt-}$ are non-vanishing.
Both of them have positive 6d chirality (\ref{6dchirality}),
and they generate
six-dimensional
${\cal N}=(1,0)$
supersymmetry.

Let us consider the condition $\delta\lambda=0$ first.
The component of $\delta\lambda$ depending on $\epsilon_{\uparrow+\wt +}$ is
\begin{align}
0=\delta\lambda^\oplus_{\downarrow-\wt +}
 &= P^{\oplus\oplus}_{z^*}\epsilon_{\uparrow+\wt +}.
\end{align}
For this to hold for $\epsilon_{\uparrow+\wt +}\neq0$,
$P^{\oplus\oplus}_{z^*}$ must vanish.
This is the same as (\ref{pzpp}) in Section \ref{7brane_in_10dim},
and the solution is given by $L=K(\tau)$ where $K(\tau)$ is defined
in (\ref{ktaudef}) with a holomorphic function $\tau(z)$.

We can also obtain similar condition for $\wt L$ form $\delta\wt\lambda=0$.
The components of $\delta\wt\lambda$ depending on $\epsilon_{\uparrow+\wt +}$ are 
$\delta\wt\lambda^{\wt i}_{\downarrow +\wt\pm}$, and we obtain the following Killing spinor equations.
\begin{align}
0=\delta\wt\lambda^{\wt i}_{\downarrow+\wt +} 
 &= \frac{i}{\sqrt{2}}\wt P^{\wt i\wt 3}_{z^*}\epsilon_{\uparrow+\wt +},\label{dchi1}\\
0=\delta\wt\lambda^{\wt i}_{\downarrow+\wt -}
 &= i\wt P^{\wt i\wt\oplus}_{z^*}\epsilon_{\uparrow+\wt +}.\label{dchi2}
\end{align}
The equation (\ref{dchi1}) requires $\wt P^{\wt i\wt3}=0$, and combining this with
(\ref{qi3}) we conclude that $L$ is essentially $SL(2)$ element.
Namely, in an appropriate choice of gauge it is given by
\begin{align}
 \wt L^{\wt\alpha}{}_{\wt i} = \wt L_0
\begin{pmatrix}
 K(\wt\tau) & 0\\
 0 & 1 
\end{pmatrix} \quad 
(\wt\tau\equiv\wt\tau_1+i\wt\tau_2\in H_+)
\end{align}
where $\wt L_0\in SL(3,\mathbb{R})$ is a constant matrix.
The condition
$\wt P^{i\wt\oplus}_{z^*}=0$
obtained from
(\ref{dchi2}) requires the function $\wt\tau$ be a holomorphic function of $z$.

Finally, we can determine the function $g$ by using $\delta\psi_m=D_m\epsilon=0$.
For this equation to hold for $\epsilon_{\uparrow +\wt+}\neq0$, the sum of three connections
$\omega$, $Q$ and $\wt Q$ must be pure gauge, and we can take the gauge
in which
\begin{align}
 \omega_{m67} + Q_{m12} + \wt Q_{m\wt 1\wt 2} =0.\label{omegaQQtilde}
\end{align}
This gives the differential equation
\begin{align}
\frac{i}{g}\partial_z g
 &= \frac{i}{2\tau_2}\partial_z\tau_2 + \frac{i}{2\wt\tau_2}\partial_z\wt\tau_2,
\end{align}
which is solved by
\begin{align}
 g = c\sqrt{\tau_2\wt\tau_2},
\end{align}
where $c$ is a positive real constant, which can be absorbed by the
coordinate change $cx^m\rightarrow x^m$.

The solution is summarized as follows.
\begin{align}
 L^\alpha{}_i &=K(\tau),\quad\tau=\tau_1+i\tau_2,\\
 \wt L^{\wt\alpha}{}_{\wt i} &= \wt L_0
\begin{pmatrix}
 K(\wt\tau) & 0\\
 0 & 1 
\end{pmatrix},
\quad
\wt L_0\in SL(3,\mathbb{R}),\quad
 \wt\tau=\wt\tau_1+i\wt\tau_2.
\end{align}
\begin{align}
ds^2=\eta_{\mu\nu}dx^\mu dx^\nu+\tau_2\wt\tau_2dx^m dx^m.
\end{align}
This is the general form of $1/4$ BPS solutions.
A solution is specified by two holomorphic functions
$\tau(z)$ and $\wt\tau(z)$ and constant $\wt L_0\in SL(3,\RR)$.

$1/2$ BPS solutions are realized as special cases of this solution.
Let us consider the case in which the supersymmetries associated with
$\epsilon_{\uparrow-\wt +}$ and its conjugate $\epsilon_{\downarrow+\wt -}$ are
also preserved
in addition to
$\epsilon_{\uparrow+\wt +}$ and its conjugate $\epsilon_{\downarrow-\wt -}$.
(\ref{6dchirality})
shows that
$\epsilon_{\uparrow-\wt +}$ and  $\epsilon_{\downarrow+\wt -}$
have negative 6d chirality and
we have ${\cal N}=(1,1)$ supersymmetry in this case.
The Killing spinor equations including $\epsilon_{\uparrow-\wt +}$ are
\begin{align}
0=\delta\lambda^\ominus_{\downarrow+\wt +}
 &= P^{\ominus\ominus}_{z^*}\epsilon_{\uparrow-\wt +}\label{dlmpp},\\
0=\delta\wt\lambda^{\wt i}_{\downarrow-\wt +} 
 &=\frac{i}{\sqrt{2}}\wt P^{\wt i\wt 3}_{z^*}\epsilon_{\uparrow-\wt +},\\
0=\delta\wt\lambda^{\wt i}_{\downarrow-\wt -} 
 &=i\wt P^{\wt i\wt\oplus}_{z^*}\epsilon_{\uparrow-\wt +}\label{dcppm},\\
0=\delta\psi_{m,\uparrow-\wt +}
&=D_m\epsilon_{\uparrow-\wt +}.
\end{align}
We have additional condition
$P^{\bm{--}}_{z^*}=0$ from (\ref{dlmpp}),
and this requires $\tau$
to be anti-holomorphic.
This means $\tau$ must be a constant.
Then the other equations hold.

There is another type of $1/2$ BPS solutions
with $\epsilon_{\uparrow+\wt -},\epsilon_{\downarrow-\wt +}\neq0$.
(\ref{6dchirality}) shows that
these components have positive 6d chirality,
and we obtain $\N=(2,0)$ supersymmetry in six dimensions.
The Killing spinor equations including $\epsilon_{\uparrow+\wt -}$ are
\begin{align}
0=\delta\lambda^\oplus_{\downarrow-\wt -}
 &= P^{\oplus\oplus}_{z^*}\epsilon_{\uparrow+\wt -},\\
0=\delta\wt\lambda^{\wt i}_{\downarrow+\wt -} 
 &= \frac{i}{\sqrt{2}}\wt P^{\wt i\wt 3}_{z^*}\epsilon_{\uparrow+\wt -},\\
0=\delta\wt\lambda^{\wt i}_{\downarrow+\wt +} 
 &=\wt P^{\wt i\wt\ominus}_{z^*}\epsilon_{\uparrow+\wt -},
\label{eq258}\\
0=\delta\psi_{m,\uparrow+\wt -}
&=D_m\epsilon_{\uparrow+\wt -}.
\end{align}
(\ref{eq258}) gives new condition
$\wt P^{\wt i\wt\ominus}_{z^*}=0$, and this means $\wt\tau$ is a constant.
Then the other conditions are satisfied.

Finally, let us consider the case with
$\epsilon_{\downarrow +\wt+}$ and $\epsilon_{\uparrow -\wt-}$ are non-vanishing.
The Killing spinor equations including $\epsilon_{\downarrow+\wt +}$ are
\begin{align}
0=\delta\lambda^\oplus_{\uparrow-\wt +} 
&= P^{\oplus\oplus}_{z}\epsilon_{\downarrow+\wt +},\\
0=\delta\wt\lambda^{\wt i}_{\uparrow+\wt +} 
&= \frac{i}{\sqrt{2}}\wt P^{\wt i\wt 3}_{z}\epsilon_{\downarrow+\wt +}, \\
0=\delta\wt\lambda^{\wt i}_{\uparrow+\wt -} 
&= i\wt P^{\wt i\wt\oplus}_{z}\epsilon_{\downarrow+\wt +},\\
0=\delta\psi_m
&=D_m\epsilon_{\downarrow+\wt +}.
\end{align}
The first gives the additional condition
$P^{\oplus\oplus}_{z}=0$,
which requires $\tau$ to be a constant,
and
the third gives
$\wt P^{\wt i\wt\oplus}_{z}=0$, and this means constant $\wt\tau$.
Then, the solution becomes trivial flat solution, and all supersymmetries are preserved.

We summarize non-trivial BPS solutions in Table \ref{tbl5}.
\begin{table}[htb]
\caption{The non-trivial 5-brane solutions in $D=8$ supergravity and world volume supersymmetries.}\label{tbl5}
\centering
\begin{tabular}{cccc}
\hline
\hline
          & ${\cal N}=(1,0)$ & ${\cal N}=(1,1)$ & ${\cal N}=(2,0)$ \\
\hline
$\tau$    & holomorphic      & constant         & holomorphic      \\
$\wt\tau$ & holomorphic      & holomorphic      & constant         \\
\hline
\end{tabular}
\end{table}

As is shown in Table \ref{tbl5}
two holomorphic functions $\tau$ and $\wt\tau$ correspond to
two types of branes.
Namely, singularities of $\tau$ and $\wt\tau$ give $5$-branes
with ${\cal N}=(2,0)$ and ${\cal N}=(1,1)$ supersymmetry, respectively.
$1/4$ BPS solutions with ${\cal N}=(1,0)$ supersymmetry is regarded as
simple superposition of two types of branes.

The most general $1/4$ BPS solution are embedded in $SL(2)\times SL(2)\subset SL(2)\times SL(3)$.
These two $SL(2)$ factors are manifest in the type IIB frame.
Namely, the $SL(2)$ factor which is a subgroup of $SL(3)$
can be associated with the axio-dilaton field in type IIB theory,
and the other $SL(2)$ is associated with the internal space $T^2$.
From the viewpoint of F-theory
the $1/4$ BPS solution can be regarded as a compactification
of the F-theory in a Calabi-Yau realized as $T^4$ fibration over $\CC$.

\subsection{$D=7$}
The $7$-dimensional maximal supergravity
has the field contents in Table \ref{7dfields} \cite{Sezgin:1982gi,Pernici:1984xx}.
\begin{table}[htb]
\caption{The field contents of $7$-dimensional maximal supergravity are shown.
Anti-symmetric tensor fields are omitted.}\label{7dfields}
\centering
\begin{tabular}{ccl}
\hline
& $SO(5)_R$ & \\
\hline
$e_M^{\wh M}$ & $1$ & vielbein \\
$L^\alpha{}_i$ & $5$ & scalars \\
$\psi_M$ & $4$ & gravitino \\
$\lambda_i$ & $16$ &  dilatino, $\rho_i\lambda_i=0$ \\
\hline
\end{tabular}
\end{table}

The scalar manifold of 7d maximal supergravity is $SL(5)/SO(5)$.
There is no duality frame which manifests whole of
the duality group $SL(5,\ZZ)$ and the R-symmetry group $SO(5)_R$.
When we regard the system as the $T^4$ compactification
of M-theory $SL(4)/SO(4)$ becomes manifest,
while $T^3$ compactification of type IIB theory manifests
$SL(3)/SO(3)\times SL(2)/SO(2)$.
Combining these we obtain the full symmetry.

The relevant part of the Lagrangian is
\begin{align}
{\cal L}
&=\frac{e}{2}R
-\frac{e}{2}(\ol\psi_M\Gamma^{MNP}D_N\psi_P)
\nonumber\\&
-\frac{e}{2}P_{Mij}P^{Mij}
-\frac{e}{2}(\ol\lambda^i\Gamma^M D_M\lambda_i)
+\frac{e}{2}(\ol\psi_M\Gamma^N\Gamma^M\rho^i\lambda^j)P_{Nij},
\end{align}
and the supersymmetry transformation rules for fermions are
\begin{align}
\delta\psi_M
&=D_M\epsilon,\nonumber\\
\delta\lambda_i
&=\frac{1}{2}P_{Mij}\Gamma^M\rho^j\epsilon.
\end{align}
The transformation parameter $\epsilon$ belongs to the
spinor representation of $H=SO(5)$, and
the covariant derivative $D_M\epsilon$ includes the connection $Q_{Mij}$.

We are interested in codimension $2$-brane solutions and
we use $x^\mu$ ($\mu=0,1,\ldots,4$) and $x^m$ ($m=5,6$) for longitudinal and transverse coordinates, respectively.
By assuming the Poincare invariance
in the five dimensions parallel to the brane,
we take the following ansatz.
\begin{align}
L=L(x^m),\quad
e^{\wh\mu}=\delta^{\wh\mu}_\mu dx^\mu,\quad
e^a=g(x^m)\delta^a_mdx^m,\quad
\epsilon=\epsilon(x^m).
\end{align}

Let us first consider the case with minimum number of unbroken supersymmetries.
The supersymmetry parameter $\epsilon$ belongs to
the  $\bm4$ of $SO(5)_R$ symmetry,
and in the minimum case
we have only one non-vanishing component.
Then the R-symmetry is broken to $SU(2)\times U(1)\subset SO(5)_R$.

It is convenient to consider the intermediate subgroup
$SU(2)_l\times SU(2)_r\sim SO(4)\subset SO(5)_R$.
The parameter $\epsilon$ is decomposed into four irreducible representation
$(2,1)_{\pm\frac{1}{2}}$ and $(1,2)_{\pm\frac{1}{2}}$
of $SU(2)_l\times SU(2)_r\times SO(2)_{56}$.
($SO(2)_{56}$ is the local Lorentz symmetry in the transverse space.)
We denote them as
\begin{align}
\epsilon\rightarrow\{\epsilon_{s,a},\epsilon_{s,\dot a}\},
\end{align}
where $s=\uparrow,\downarrow$ represent the $SO(2)_{56}$ charges
and $a$ and $\dot a$ are indices for $SU(2)_l$ and $SU(2)_r$, respectively.
The fields $P$ and $Q$ are decomposed as
\begin{align}
Q_{ij}\rightarrow\{Q_{a\dot a},Q_{(ab)},Q_{(\dot a\dot b)}\},\quad
P_{ij}\rightarrow
\{P,P_{a\dot b},P_{(ab)(\dot a\dot b)}\},
\end{align}
and the dilatino $\lambda$ as
\begin{align}
\lambda\rightarrow
\{\lambda_{s a},
\lambda_{s\dot a},
\lambda_{s(ab)\dot a},
\lambda_{sa(\dot a\dot b)}\},
\end{align}
where a pair of indices in parenthesis are symmetric.
If we choose $\epsilon_{\dot1}$ as the component for the unbroken supersymmetry,
$SU(2)_r$ is broken to $U(1)_r$.
The connection $Q$ should take its value in $SU(2)_l\times U(1)_r$.
Namely, $Q_{a\dot a}=Q_{\dot1\dot1}=Q_{\dot2\dot2}=0$
and the only non-vanishing components are
\begin{align}
Q_{(ab)},\quad
Q_{\dot1\dot2}.
\label{7dq}
\end{align}

$\epsilon_{\uparrow\dot a}$ appear in the supersymmetry transformation as
\begin{align}
\delta\lambda_{\downarrow a}&=P_{z^*a\dot b}\epsilon_{\uparrow}^{\dot b},\label{7dlambda1}\\
\delta\lambda_{\downarrow\dot a}&=P_{z^*}\epsilon_{\uparrow\dot a},\label{7dlambda2}\\
\delta\lambda_{\downarrow(ab)\dot a}&=P_{z^*(ab)(\dot a\dot b)}\epsilon_{\uparrow}^{\dot b},\label{7dlambda3}\\
\delta\lambda_{\downarrow a(\dot a\dot b)}&=P_{z^*a(\dot a}\epsilon_{\uparrow\dot b)}.\label{7dlambda4}
\end{align}
(We omitted the numerical coefficients that are not important here.)
If we require $\delta\lambda=0$ for $\epsilon_{\uparrow\dot1}\neq 0$,
we obtain $P_{z^*}=P_{z^*a\dot a}=P_{z^*(ab)(\dot a\dot2)}=0$
and the only non-vanishing components of $P_{z^*}$ are
\begin{align}
P_{z^*(ab)(\dot1\dot1)}.
\label{7dp}
\end{align}
(We also have similar conditions for $P_z$ from the equations containing 
$\epsilon_{\downarrow\dot 2}\sim(\epsilon_{\uparrow\dot 1})^*.$)
(\ref{7dq}) and (\ref{7dp}) show that non-vanishing components of
$P$ and $Q$ are associated with a subgroup $SO(4)\subset SO(5)_R$.
As we mentioned above $SO(4)$ subgroup
of $SO(5)_R$ can be realized geometrically if we regard the theory as
$T^4$ compactification of M-theory.

We want to give the scalar fields $L$ such that $P$ and $Q$ have
only non-vanishing components (\ref{7dp}) and (\ref{7dq}).
Unfortunately, we have not obtained the answer.
To simplify the problem,
let us consider a restricted case with
$P_{(12)}=0$.
Then $F^{(Q)}=P\wedge P$ takes value in the Cartan part of $SU(2)\times U(1)$,
and we can take the gauge such that
$Q_{(11)}=Q_{(22)}=0$,
then non-vanishing components are
\begin{align}
Q_{(12)},\quad
Q_{(\dot 1\dot 2)},\quad
P_{(11)(\dot1\dot1)},\quad
P_{(22)(\dot1\dot1)}.
\end{align}
In this case, with an appropriate real basis,
the $5\times5$ matrices $P$ and $Q$ are
block diagonal matrices in the following form.
\begin{align}
Q=\left(\begin{array}{ccc}
Q' & & \\
& Q'' & \\
&& 0
\end{array}\right),\quad
P=\left(\begin{array}{ccc}
P' & & \\
& P'' & \\
&& 0
\end{array}\right).
\end{align}
Therefore, the solution reduces to the superposition of two copies of
solutions for
$SL(2,\RR)/SO(2)$ scalar manifold.
In the same way as in higher dimensions,
each $SL(2,\RR)$ part can be expressed in terms of a holomorphic function.
Let the two holomorphic functions be $\tau'$ and $\tau''$.
The solution is given by
\begin{align}
L^\alpha{}_i&=\left(\begin{array}{ccc}
K(\tau') \\
& K(\tau'') \\
&&0
\end{array}\right),\\
ds^2&=\eta_{\mu\nu}dx^\mu dx^\nu+\tau'_2\tau''_2dx^m dx^m.
\end{align}

As a special case of this $1/4$ BPS solution we can realize $1/2$ BPS solution.
Let us consider the cases there is another Killing spinor in addition to $\epsilon_{\dot1}$.
There are two cases.

First, let us consider the case that $\epsilon_1$ is also a Killing spinor.
In this case, from
\begin{align}
0=\delta\lambda_{\downarrow(\dot a\dot b)a}=P_{z^*(ab)(\dot a\dot b)}\epsilon_{\uparrow}^b
\end{align}
we obtain $P_{z^*(a2)(\dot a\dot b)}=0$.
Then the only non-vanishing component of $P_{z^*}$ is
$P_{z^*(11)(\dot1\dot1)}$.
In this case, just like the case of $1/2$ BPS solution in 8d,
we can show that one of $\tau'$ and $\tau''$ must be $z$-independent constant.

If two Killing spinor have the same $SO(4)_R$ chirality,
the equation (\ref{7dlambda3}) require $P_{z^*(ab)(\dot a\dot b)}=0$.
Namely, all components of $P$ vanish.
Because $F^{(Q)}=P\wedge P=0$
we can choose a gauge with $Q=0$.
Therefore, the solution is trivial.

\section{Conclusions}

In this paper we investigated codimension-2 BPS solutions
in maximal supergravities in $9$, $8$, and $7$ dimensions.
We assumed the Poincare invariance along branes and
vanishing of various gauge fields.

The scalar manifold of $9$d maximal supergravity is $(SL(2,\RR)/SO(2))\times\RR$.
In a BPS solution the scalar field associated with the factor $\RR$ must be constant
and play no role.
Therefore, the solutions are essentially the same as those of type IIB supergravity in $10$d,
and simply interpreted as the double dimensional reduction of type IIB $7$-branes.

In $8$d, the scalar manifold consists of two factors $SL(2,\RR)/SO(2)$ and $SL(3,\RR)/SO(3)$.
Killing spinor equations associated with these factors decouple,
and we can solve them one by one.
From the $SL(2,\RR)/SO(2)$ part we obtain $1/2$ BPS branes on which six-dimensional ${\cal N}=(2,0)$ supersymmetry is
realized
while from the $SL(3,\RR)/SO(3)$ part we obtain $1/2$ BPS solutions on which six-dimensional ${\cal N}=(1,1)$ supersymmetry
is realized.
The latter is always embedded in $SL(2,\RR)/SO(2)\subset SL(3,\RR)/SO(3)$.
These two types of solutions
have essentially the same structure
as the $7$-brane solution in $10$d.
Namely, each type of classical solution is specified by a holomorphic function
and singularities of the function give branes.
We also found $1/4$ BPS solutions,
which are specified by two holomorphic functions and are regarded as
simple superposition of two types of $1/2$ BPS solutions.
If we regard the $8$d supergravity as the $T^2$ compactification of type IIB theory,
the two copies of $SL(2,\RR)/SO(2)$ are associated with the complex moduli of the
torus and the axio-dilaton field in type IIB theory,
and both are geometrically realized in F-theory.

In $7$d, a generic BPS solution is $1/4$-BPS.
We showed that such solution can be embedded in $SL(4,\RR)/SO(4)\subset SL(5,\RR)/SO(5)$.
This means the solution can be realized as a geometric compactification
of M-theory.
We could not solve the Killing spinor solutions in the general situation.
We introduced one additional restriction to simplify the problem, and then the solution is
factorized into two copies of solutions associated with $SL(2,\RR)/SO(2)$.
Again, similarly to the $8$d case, the solution can be regarded as
a simple superposition of
two $1/2$ BPS solutions.

Although our original motivation of this work was to find essentially 
new BPS branes,
all solutions we found have simple geometric realization in M- or F-theory.

\section*{Acknowledgments}
We would like to thank Tetsuji Kimura for valuable discussions.
The work of Y.I. was partially supported by Grand-in-Aid for Scientific Research (C) (No.15K05044),
Ministry of Education, Science and Culture, Japan.

\appendix
\section{Notes on our notation}

We denote the dirac matrices for $SO(2)_R$ group by $\rho_i$ ($i=1,2$).
We use the representation
\begin{align}
\rho_1=\sigma_x,\quad
\rho_2=\sigma_y,
\end{align}where $\sigma_{x,y,z}$ are the Pauli matrices.
We use lower indices $a,b,\ldots$ for two-component spinors,
and thus the matrices acting on them have lower and upper indices like $(\rho_i)_a{}^b$.
For components of spinors we define the $SO(2)_R$ charge as
eigenvalues of the generator $(-i/2)\rho_{12}$.
This means the upper and the lower components of spinors carry the charges $+1/2$ and $-1/2$, respectively.
To specify these components of a spinor $\chi$ we use the notation $\chi_+$ and $\chi_-$, respectively.

For the analysis of the Killing spinor equations it is convenient to the
complex basis for vectors.
For example, for a vector $v^i$ ($i=1,2$) we define
$v^\oplus=v_\ominus=\frac{1}{\sqrt2}(v^1+iv^2)$ and
$v^\ominus=v_\oplus=\frac{1}{\sqrt2}(v^1-iv^2)$.
For $\rho^i$ we have
\begin{align}
\rho^\oplus=\rho_\ominus=\left(\begin{array}{cc}
0 & \sqrt2 \\
0 & 0
\end{array}\right),\quad
\rho^\ominus=\rho_\oplus=\left(\begin{array}{cc}
0 & 0 \\
\sqrt2 & 0\
\end{array}\right).
\end{align}
With this representation the lower index $\oplus$ and upper index $\ominus$ carry $SO(2)_R$ charge $+1$,
while the lower $\ominus$ and upper $\oplus$ carry $SO(2)_R$ charge $-1$.
This is checked by looking at the non-vanishing components of $\rho^i$.
For example the non-vanishing component of $\rho_\oplus$ is $(\rho_\oplus)_-{}^+$,
and the total charge of this component must be zero.
The statement above about $SO(2)_R$ charge is consistent to this.

For the local rotation symmetry in the transverse space to branes
we use up and down for the $SO(2)$ charge (spin) $\pm1/2$.
With the choice of the Dirac matrices,
the lower indices $z$ and $z^*$ carry spin $+1$ and $-1$,
respectively.

In $8$d we also deal with $SO(3)_R$ symmetry.
The notation is basically the same as the $SO(2)_R$ case except we put tildes on variables and indices
for distinction from $SO(2)_R$ objects.
We specify components of spinors by eigenvalues of the Cartan generator $(-i/2)\wt\rho_{\wt 1\wt 2}$.
$\wt\chi_{\wt\pm}$ carry the charge $\pm1/2$,
and a vector $\wt v$ has three components
$\wt v^{\wt\ominus}=\wt v_{\wt\oplus}$,
$\wt v^{\wt\oplus}=\wt v_{\wt\ominus}$, and
$\wt v^3=\wt v_3$
that carry the Cartan charge $+1$, $-1$, and $0$, respectively.

In the following we give relations of fields in this paper and those in references.
We will not give detailed explanations for normalization of fields, spinor conventions,
etc., because they are not
important in our analysis of the Killing spinor equations.
We focus on giving rough correspondence between fields used in
this paper and those in references.

\paragraph{$D=10$}
Ten dimensional supergravity is given in \cite{SchwarzIIB}.
The global symmetry $SL(2,\RR)$ is isomorphic to $SU(1,1)$.
In \cite{SchwarzIIB} the scalar fields are expressed as the matrix $V_\pm^a$,
which is defined with a complex basis natural for $SU(1,1)$.
The real matrix $L^\alpha{}_i$ used in this paper is
related with $V$ by
\begin{align}
\left(\begin{array}{cc}
V^1{}_- & V^1{}_+ \\
V^2{}_- & V^2{}_+
\end{array}\right)
=
U\left(\begin{array}{cc}
L^1{}_1 & L^1{}_2 \\
L^2{}_1 & L^2{}_2
\end{array}\right)U^\dagger,\quad
U=\frac{1}{\sqrt{2}}\left(\begin{array}{cc}
1 & -i \\
1 & i
\end{array}\right).
\end{align}

The dilatino field $\lambda$ in \cite{SchwarzIIB}
is
defined as the field with $U(1)_R$ charge $\pm3/2$,
while we denote this as a field with vector and spinor indices.
They are related by
\begin{align}
\lambda\sim \lambda_+^\ominus,\quad
\ol\lambda\sim \lambda_-^\oplus.
\end{align}
Due to the $\rho$-traceless condition $\lambda^\oplus_+=\lambda^\ominus_-=0$.

\paragraph{$D=9$}
The $9$-dimensional maximal supergravity is given in \cite{Nishino:2002zi}.
Two dilatino fields in \cite{Nishino:2002zi}
are renames as $\lambda_i$ and $\wt\lambda$ to
to match the fields in the other dimensions.
$SO(2)$ Dirac matrices
are denoted by $\tau_i$ in \cite{Nishino:2002zi}
while we use $\rho_i$ for them.

\paragraph{$D=8$}
The 8-dimensional maximal supergravity is given in \cite{Salam:1984ft}.
The dilatino field $\chi_i$ in \cite{Salam:1984ft} does not satisfy the $\rho$-traceless condition,
and we decompose it into the traceless part $\wt\lambda_i$ and the trace part $\lambda_I$.
To make the $SL(2,\RR)/SO(2)$ structure manifest
we combine the scalar fields $\phi$ and $B$ in \cite{Salam:1984ft}
into the matrix $L^A{}_I$.
With the gauge choice like (\ref{tendimL})
these are related by $L=K(\tau)$ with $\tau=-2B+ie^{2\phi}$.

\paragraph{$D=7$}
The $7$-dimensional maximal supergravity is given in \cite{Pernici:1984xx}.
In the reference the scalar matrix is denoted by $\Pi$ instead of $L$.
Notation for other fields is similar to ours.

\end{document}